\documentstyle[12pt]{article}
\normalbaselines
\catcode `\@=11
\@addtoreset{equation}{section}
\def\section{\@startsection {section}{1}{\z@}
            {-3.5ex plus -1ex minus -.2ex}{2.3ex plus .2ex}{\normalsize\bf}}
\def\subsection{\@startsection{subsection}{2}{\z@}
            {-3.25ex plus -1ex minus -.2ex}{1.5ex plus .2ex}{\normalsize\sl}}
\def\thebibliography#1{\section*{References\markboth{REFERENCES}{REFERENCES}}
 \list{[\arabic{enumi}]}{\settowidth\labelwidth{[#1]}\leftmargin\labelwidth
 \advance\leftmargin\labelsep\usecounter{enumi}}
 \def\newblock{\hskip .11em plus .33em minus -.07em}\sloppy
 \sfcode`\.1000\relax}
 
\catcode `\@=12

\skip\footins=14pt % was 12
\begin{document}
\noindent
\hfil January 2001
\break
\hrule

\vspace*{1.8cm}
\noindent
{\bf DIMENSIONAL REDUCTION BY A TWO-FORM}\\[1mm]
{\bf (another alternative to compactification)}\vspace{1.0cm}\\
\noindent
\hspace*{1in}
\begin{minipage}{13cm}
\makebox[3mm]P. Maraner\vspace{0.3cm}\\
\makebox[3mm]\ \\%via Capri 13\\
\makebox[3mm]\ \\%39100 Bolzano, Italy\\
\end{minipage}

\vspace*{0.3cm}

\begin{abstract}
\noindent
It is shown that the local coupling of a higher dimensional graviton 
to a closed degenerate two-form produces dimensional reduction by
 spontaneous breakdown of extra-dimensional translational symmetry. 
Four dimensional Poincar\'e invariance emerges as residual symmetry. 
As a specific example, a six dimensional geometry coupled to a closed 
rank 2 two-form yields the `ground state'
$$ds^2={\rm e}^{-|\xi|^2/4l^2}\eta_{\mu\nu}dx^\mu dx^\nu+\delta_{ij}
d\xi^i d\xi^j$$
with $l$ a fundamental length scale. At low energies, space-time 
reduces to four observable dimensions and general relativistic gravity 
is reproduced.
\end{abstract}

\noindent
There has been recently a renewed interest in higher dimensional geometry
as an intermediate step in between today's high energies field theories and
a foretold final theory. Besides the search for phenomenological implications, 
one of the hottest problems remains that of reproducing the effective four 
dimensional nature of space-time at low energies scales. 
To one side the traditional hypothesis of compact extra dimensions has been 
fully reconsidered \cite{AADD}. To the other, a remarkable alternative has 
been put forward by showing that dimensional reduction take place without 
compactification in a set-up with a 3-brane embedded in the higher dimensional
space \cite{Randall-Sundrum}. In both approaches dimensional reduction is 
produced by global topological assumptions --compactness of extra-dimensions 
in the first case, 3-brane boundary conditions in the second one. \\
 In this letter we explore an alternative based on the local interaction 
with a closed degenerate two-form. The specific model we consider is a six 
dimensional space-time $(M^6, {\rm g}_{IJ})$ on which a closed rank $2$ 
two-form ${\rm B}_{IJ}$ is defined\footnote{Capital Latin indices run over
$I=0,1,...,5$}. No physical hypothesis are made on the nature of 
${\rm B}_{IJ}$. In solving generalized Einstein's equations it is found that 
translational invariance is spontaneously broken and four dimensional 
Poincar\'e invariance emerges as the residual symmetry of the geometric 
`ground state'. 
The corresponding metric tensor reproduces --up to the functional form of the 
wrap factor and two instead of one extra dimensions-- the one recently 
investigated by L.\ Randall and R.\ Sundrum \cite{Randall-Sundrum}. 
 On low energies scales space-time dynamically reduces to four effective 
dimensions and the ordinary general relativistic picture is recovered.

\vskip0.3cm
 {\it Field Equations.} The reason for such a behavior grounds in the 
peculiar properties of the local coupling with a closed degenerate two-form. 
In looking for field equations we first observe
that skew-symmetry and closure condition $(d{\rm B})_{IJK}=0$ 
make ${\rm B}_{IJ}$ very similar to an ordinary electromagnetic field. 
In particular, it is possible to introduce a vector potential ${\rm A}_I$
which six dimensional curl equals ${\rm B}_{IJ}$
\begin{equation}
(d{\rm A})_{IJ}={\rm B}_{IJ}
\end{equation}
As in ordinary electromagnetism ${\rm A}_I$ is determined up to
the gradient of an arbitrary function only. The transformation
${\rm A}_I\rightarrow {\rm A}_I +\partial_I\chi$ 
leaves ${\rm B}_{IJ}$ unchanged. By this analogy, we are lead to 
include in the Lagrangian density of the theory
a term proportional to ${\rm B}_{IJ}{\rm B}^{IJ}$ 
--besides the scalar curvature ${\rm R}$ and the cosmological term.
Other contributions containing ${\rm B}_{IJ}$ or its covariant
derivatives may be assumed higher order.
At this point however, we should recall that the four dimensional 
electromagnetic analogy is purely formal. In a higher dimensional
space-time the one-form ${\rm A}_I$ still can be coupled to geometry by 
giving a `charge' to the graviton. 
By extending a well established strategy, 
${\rm g}_{IJ}$ is promoted to a complex field\footnote{As for an
ordinary charged scalar or fermion field, the complex nature of the 
metric tensor has nothing to do with its space-time transformation 
properties.} and everywhere in field equations ordinary derivatives 
$\partial_I$ are replaced by gauge covariant ones
\begin{equation}
\partial_I-il^{-2}{\rm A}_I
\end{equation}
where (by choosing ${\rm B}_{IJ}$ dimensionless) $l$ carries
the dimension of a length. 
Clearly, by a redefinition of vector potential ${\rm A}_I\rightarrow
{\rm A}_I+\partial_I\chi$ the metric field is assumed to transform
according to ${\rm g}_{IJ}\rightarrow {\rm e}^{i\chi/l^2}{\rm g}_{IJ}$.\\
 In this way we come to the following generalization of Einstein's field 
equations 
\begin{equation}
{{\rm R}\hskip-0.3cm/}\hskip0.14cm_{IJ}
-{1\over2}{{\rm R}\hskip-0.3cm/}\hskip0.14cm {\rm g}_{IJ}=
 {\sl\Lambda}\!\  {\rm g}_{IJ}
-K\left({\rm B}_I^{\ K}{\rm B}_{JK}
        -{1\over4}{\rm B}_{KL}{\rm B}^{KL}{\rm g}_{IJ}\right)
\label{fe}
\end{equation}
where ${\sl\Lambda}$ and $K$ are constants, ${\rm g}_{IJ}$ is now a 
complex field and Dirac's slash notation has been borrowed to denote 
quantities in which ordinary derivatives are replaced by gauge covariant ones. 
 
\vskip0.3cm
 {\it Adapted Coordinates.} A remarkable property of Eqns.(\ref{fe})
is that for no choice of ${\sl\Lambda}$, $K$ and no ${\rm A}_I$ gauge,
the flat space-time Minkowskian metric $\eta_{IJ}$ is a solution. To
prove this and solve field equations it is very convenient to 
work in an adapted coordinates system. Given the closure condition 
$(d{\rm B})_{IJK}=0$, a classical theorem of Darboux ensures the 
possibility of globally finding coordinates in such a way that 
\begin{equation}
B_{IJ}=
\left(
\begin{array}{cccccc}
 0 & 0 &   &   &   &   \cr
 0 & 0 &   &   &   &   \cr
   &   & 0 & 0 &   &   \cr
   &   & 0 & 0 &   &   \cr
   &   &   &   & 0 & 1 \cr
   &   &   &   &-1 & 0
\end{array}
\right)
\nonumber
\end{equation}  
Further assuming ${\rm B}_{IJ}$ non null directions to be space-like,
Darboux coordinates are denoted by ${\rm x}^I=(x^\mu,\xi^i)$ with
$\mu=0,1,2,3$ and $i=4,5$. In such coordinates frames ${\rm A}_I$
can be chosen in the form ${\rm A}_I=(0,A_i)$ with $A_i$ the vector
potential of a two-dimensional homogeneous magnetic field
--e.g.\ in the symmetric gauge $A_i=\left(\xi^5/2,-\xi^4/2\right)$. 

\vskip0.3cm
 {\it Four Effective Dimensions.} Assuming the geometric `ground state'
to be compatible with four dimensional Poincar\'e invariance, we are lead
to the following ansatz for the six dimensional line element
\cite{Randall-Sundrum}
\begin{equation}
ds^2=\phi(\xi)\eta_{\mu\nu}dx^\mu dx^\nu+ \delta_{ij}d\xi^i d\xi^j
\label{ansatz}
\end{equation}
With this choice field equations take the form
\begin{eqnarray}
&&\hskip-1.2cm
{1\over l^2} \left( {\partial\hskip-0.22cm/}_4{\partial\hskip-0.22cm/}_4
+{\partial\hskip-0.22cm/}_5{\partial\hskip-0.22cm/}_5\right)\phi
\ \!\eta_{\mu\nu}=
{1\over3}\left(2{\sl\Lambda}-K\right)\phi\ \!\eta_{\mu\nu} \label{fe1}\\[2mm]
&&\hskip-1.2cm
{1\over l^2}\!\ {\partial\hskip-0.22cm/}_5{\partial\hskip-0.22cm/}_5\phi 
-{i\over 2l^3}\!\ A_4\!\ {\partial\hskip-0.22cm/}_4\phi 
+{i\over 2l^3}\!\ A_5\!\ {\partial\hskip-0.22cm/}_5\phi
+{3\over4l^2}{\left({\partial\hskip-0.22cm/}_4\phi\right)^2\over\phi}
+{1\over4l^2}{\left({\partial\hskip-0.22cm/}_5\phi\right)^2\over\phi}
={K\over4}\ \!\phi \\[2mm]
&&\hskip-1.2cm
{1\over l^2} \left({\partial\hskip-0.22cm/}_4{\partial\hskip-0.22cm/}_5
+{\partial\hskip-0.22cm/}_5{\partial\hskip-0.22cm/}_4\right)\phi
+{i\over l^3}\!\ A_4\!\ {\partial\hskip-0.22cm/}_5\phi
+{i\over l^3}\!\ A_5\!\ {\partial\hskip-0.22cm/}_4\phi
-{1\over l^2}{\left({\partial\hskip-0.22cm/}_4\phi\right)
  \left({\partial\hskip-0.22cm/}_5\phi\right)\over\phi}=0\\[2mm]
&&\hskip-1.2cm
{1\over l^2}\!\ {\partial\hskip-0.22cm/}_4{\partial\hskip-0.22cm/}_4\phi 
+{i\over 2l^3}\!\ A_4\!\ {\partial\hskip-0.22cm/}_4\phi
-{i\over 2l^3}\!\ A_5\!\ {\partial\hskip-0.22cm/}_5\phi
+{1\over4l^2}{\left({\partial\hskip-0.22cm/}_4\phi\right)^2\over\phi}
+{3\over4l^2}{\left({\partial\hskip-0.22cm/}_5\phi\right)^2\over\phi}
={K\over4}\ \!\phi 
\end{eqnarray}
where  dimensionless operators 
${\partial\hskip-0.22cm/}_i=l\partial_i-il^{-1}A_i$ 
have been introduced.
${\partial\hskip-0.22cm/}_4$ 
and
${\partial\hskip-0.22cm/}_5$ 
have the form of  kinematical momenta for a charged particle 
coupled to a homogeneous magnetic field.  They close  canonical 
commutation relations
\begin{equation}
\left[{\partial\hskip-0.22cm/}_4,{\partial\hskip-0.22cm/}_5\right]=i
\end{equation} 
By this analogy eq.(\ref{fe1}) is identified with the wave equation 
of a scalar particle in a two dimensional homogeneous magnetic field. 
All solutions are easily constructed. No constant is included. 
By replacing them in the remaining equations we find --up to a normalization 
factor ${\cal N}$ and an arbitrary point $\bar\xi$ in extra directions-- 
the only solution
\begin{eqnarray}
\phi(\xi)={\cal N}{\rm e}^{-|\xi-\bar\xi|^2/4l^2}
\end{eqnarray}
with $|\xi|^2={\xi^4}^2+{\xi^5}^2$ and
corresponding to the values ${\sl\Lambda}=-5/2l^2$ and $K=-2/l^2$. 
Conventionally, we set ${\cal N}=1$ and choose $\bar\xi$ as the origin of
extra coordinates. In this way the line element relative to the geometric 
`ground state' is univocally determined as
\begin{equation}
ds^2={\rm e}^{-|\xi|^2/4l^2}\eta_{\mu\nu}dx^\mu dx^\nu 
     +\delta_{ij} d\xi^i d\xi^j
\label{estm}
\end{equation}
Translational symmetry in extra directions is spontaneously broken. At energy 
scales small compared to the one associated to the length $l$, space-time is 
effectively squeezed on four ordinary directions while the two extra ones are 
made inaccessible by the Gaussian factor ${\rm e}^{-|\xi|^2/4l^2}$.
Up to the functional form of the wrap factor $\phi(\xi)$ and 
two instead of one extra dimensions, the metric we find
resembles the one recently studied by Randall and Sundrum.
 The main difference in the two approaches is that in 
ref.\cite{Randall-Sundrum} translational invariance in extra directions
is broken from the very beginning by assuming the existence 
of the 3-brane while here it is 
spontaneously broken by assuming the local interaction with the closed 
degenerate two-form ${\rm B}_{IJ}$.  
 Apart form this, the low energy scenario associated to the metric tensor
(\ref{estm}) shares most of the characteristics emphasized by Randall and 
Sundrum. We refer to their papers for a detailed discussion.

\vskip0.3cm
 {\it Four Dimensional Gravity.}  By a slight modification we can easily 
construct solutions corresponding to a gravitational field in the effective 
four dimensional space-time. In (\ref{ansatz}), we replace the
flat four dimensional metric $\eta_{\mu\nu}$ by an arbitrary metric 
$g_{\mu\nu}(x)$
\begin{equation}
ds^2=\phi(\xi)g_{\mu\nu}(x)dx^\mu dx^\nu+ \delta_{ij} d\xi^i d\xi^j
\end{equation}
By replacing this new ansatz in field equations we are lead to
the differential problem
\begin{eqnarray}
&&\hskip-1.2cm
{1\over l^2} \left( {\partial\hskip-0.22cm/}_4{\partial\hskip-0.22cm/}_4
+{\partial\hskip-0.22cm/}_5{\partial\hskip-0.22cm/}_5\right)\phi\ \!g_{\mu\nu} 
+{2\over3}\left(R_{\mu\nu}-{1\over2}R\!\ g_{\mu\nu}\right)=
{1\over3}\left(2{\sl\Lambda}-K\right)\phi\ \!g_{\mu\nu} \\[2mm]
&&\hskip-1.2cm
{1\over l^2}\!\ {\partial\hskip-0.22cm/}_5{\partial\hskip-0.22cm/}_5\phi 
-{i\over 2l^3}\!\ A_4\!\ {\partial\hskip-0.22cm/}_4\phi 
+{i\over 2l^3}\!\ A_5\!\ {\partial\hskip-0.22cm/}_5\phi
+{3\over4l^2}{\left({\partial\hskip-0.22cm/}_4\phi\right)^2\over\phi}
+{1\over4l^2}{\left({\partial\hskip-0.22cm/}_5\phi\right)^2\over\phi}
-{1\over4}R
={K\over4}\ \!\phi \\[2mm]
&&\hskip-1.2cm
{1\over l^2} \left({\partial\hskip-0.22cm/}_4{\partial\hskip-0.22cm/}_5
+{\partial\hskip-0.22cm/}_5{\partial\hskip-0.22cm/}_4\right)\phi
+{i\over l}\!\ A_4\!\ {\partial\hskip-0.22cm/}_5\phi
+{i\over l}\!\ A_5\!\ {\partial\hskip-0.22cm/}_4\phi
-{1\over l^2}{\left({\partial\hskip-0.22cm/}_4\phi\right)
  \left({\partial\hskip-0.22cm/}_5\phi\right)\over\phi}=0\\[2mm]
&&\hskip-1.2cm
{1\over l^2}\!\ {\partial\hskip-0.22cm/}_4{\partial\hskip-0.22cm/}_4\phi 
+{i\over 2l^3}\!\ A_4\!\ {\partial\hskip-0.22cm/}_4\phi
-{i\over 2l^3}\!\ A_5\!\ {\partial\hskip-0.22cm/}_5\phi
+{1\over4l^2}{\left({\partial\hskip-0.22cm/}_4\phi\right)^2\over\phi}
+{3\over4l^2}{\left({\partial\hskip-0.22cm/}_5\phi\right)^2\over\phi}
-{1\over4}R
={K\over4}\ \!\phi 
\end{eqnarray}
where $R_{\mu\nu}$ and $R$ denote respectively Riemann tensor and
scalar curvature associated to $g_{\mu\nu}$. 
The problem exactly separates in the one we just solved for the freedom 
$\phi(\xi)$ plus a standard Einstein problem for the four dimensional 
metric $g_{\mu\nu}$ 
\begin{equation}
R_{\mu\nu}-{1\over2}R\!\ g_{\mu\nu}=0
\end{equation}
While on energy scales of order $l^{-2}$ space-time is squeezed on a $1+3$ 
pseudo-Riemannian hyper-surface, the low energy --order one-- dynamics of the 
effective four dimensional geometry is described by Einstein field equations.

\vskip0.3cm
 {\it Other Fields.} In concluding, we remark that the coupling 
of ${\rm B}_{IJ}$ to other higher dimensional `charged' fields 
--either bosonic or fermionic-- also produces their squeezing 
on the effective four dimensional space-time by a very similar 
mechanism \cite{Maraner}.

\vskip0.7cm
\noindent
{\bf Acknowledgements}: I wish to acknowledge useful conversations with
Mario Tonin and Roberto De Pietri.

\end{document}